\documentclass[twocolumn,prl,showpacs]{revtex4}
\usepackage{graphicx}% Include figure files
\usepackage{bm}% bold math

\begin{document}
%
%----------------------------------------------------------------------------

\title{Anomalous electronic correlations in ground state
momentum density of Al${_{97}}$Li${_{3}}$}

%-----------------------------------------------------------------------------

\author{J. Kwiatkowska$^1$,
B. Barbiellini$^2$, S. Kaprzyk$^{2,3}$, A. Bansil$^2$,
H. Kawata$^4$ and N. Shiotani$^4$}
\address{
$^1$H. Niewodnicza\'nski Institute of Nuclear Physics, 
Polish Academy of Sciences,\\ 
Radzikowskiego 152, 31-342 Krak\'ow, Poland\\
$^2$Department of Physics,
Northeastern University, Boston, MA 02115, USA\\
$^3$Academy of Mining and Metallurgy AGH, 
Al. Mickiewicza 30, 30-059 Krak\'ow, Poland\\
$^4$Photon Factory, High Energy Accelerator Research Organization,
Tsukuba, Ibaraki 305-0801, Japan}

\date{\today}

\begin{abstract}

We report high resolution Compton scattering measurements on an 
Al$_{97}$Li$_3$ disordered alloy single crystal for momentum transfer 
along the [100], [110] and [111] symmetry directions. The results are 
interpreted via corresponding KKR-CPA (Korringa-Kohn-Rostoker coherent 
potential approximation) first principles computations. By comparing 
spectra for Al$_{97}$Li$_3$ and Al, we show that the momentum density in 
the alloy differs significantly from the predictions of the conventional 
Fermi liquid picture and that the ground state of Al is modified 
anomalously by the addition of Li.

\end{abstract}
\maketitle

% 71.23.-k Electronic structure of disordered solids
% 71.45.Gm Exchange, correlation, dielectric and magnetic response
%functions, plasmons
% 78.70.Ck X-ray scattering
% 41.60.Ap Synchrotron radiation
% 32.80.Cy Atomic scattering, cross sections, and form factors; Compton
%scattering

The addition of Li to Al not only reduces the density below that of Al but 
also increases the elastic modulus. The resulting Al-Li alloy possesses a 
high strength to weight ratio, which makes it well suited for applications 
such as fuel efficient aircraft components \cite{martin_review}. 
Metallurgically, Al is quite 'exclusive' in that only a few percent 
impurities can be dissolved in Al in the solid solution $\alpha$ phase. 
For such technological and fundamental reasons, the electronic structure 
and bonding properties of Al-Li alloys have been the subject of 
considerable attention over the years \cite{ceder}. Here we directly probe 
changes in the electronic ground state of Al due to the presence of Li 
impurities via Compton scattering measurements. We find that the 
modifications in the momentum density of Al induced by just a few percent 
Li atoms are surprisingly large and cannot be accounted for within the 
standard Fermi liquid type model of the correlated homogeneous electron 
gas. We show however that the experimentally observed momentum density in 
Al$_{97}$Li$_3$ can be described reasonably well if correlation effects 
missing in the standard picture are modeled by promoting 3\% electrons in 
the system from $s$ to $p$ orbitals. Given the increasing current interest 
in understanding correlation effects in the inhomogeneous electron gas, 
our results indicate that Al-Li alloys present an example of a simple 
binary system which exhibits unusual correlation effects even though each 
of its two constituents is commonly thought of as being a free electron 
like metal.

Compton scattering \cite{cooper,kaplan} is one of the few spectroscopies
capable of directly probing the bulk electronic ground state in materials.
The measured double-differential cross-section, usually referred to as the
Compton profile (CP), is given by
\begin{equation}
J(p_z)=\int\int n({\bf p})dp_x dp_y~,
\label{equ:001}
\end{equation}
where $n({\bf p})$ is the ground-state electron momentum density. The 
theoretical analysis of the Compton spectra is often based on the 
expression of $n({\bf p})$ within the independent particle model 
\cite{kaplan}. When the right hand side of Eq.~\ref{equ:001} is evaluated 
using the selfconsistent band theory based electron wavefunctions in the 
local density approximation (LDA), supplemented with electron correlation 
effects in the homogeneous electron gas, one obtains a fairly 
sophisticated description of the Compton spectrum. In a randomly 
disordered alloy, the ensemble averaged momentum density can be obtained 
within the first-principles KKR-CPA framework \cite{bansil_jpcs}. We note 
two previous high resolution Compton scattering studies of Al-Li 
\cite{matsu,suortti01}. Ref.~\onlinecite{matsu} focuses on the Fermi 
surface (FS) of Al$_{97}$Li$_3$ and by applying a novel reconstruction 
technique shows the FS to be in accord with the KKR-CPA predictions. 
Ref.~\onlinecite{suortti01} attempts to examine changes in the CPs with 
alloying and concludes that within the resolution and statistics of the 
experiment, the data are consistent with free electron behavior \cite{matsu0}.

With regard to experimental details, a single crystal of Al$_{97}$Li$_3$ 
alloy was grown by Bridgman method. The starting material was an alloy 
made from high purity Al (99.999\%) and Li (99.95\%), containing excess of 
Li above the desired composition to allow for losses.  The load was kept 
under 3 atm. Ar pressure during the process of crystallization. The final 
Li content in the alloy single crystal was determined by Atomic Absorption 
Spectrometry to be 3 at.\%. The Li content was also checked independently 
using Proton Induced Gamma Emission. The crystal was oriented using Laue 
X-ray diffraction method and disc-shaped samples, 10mm in diameter and 
1.5mm thick, were cut parallel to the (100), (110), and (111) 
crystallographic planes. The CPs were measured with the high 
resolution Compton spectrometer installed at the KEK-AR synchrotron \cite{sakurai92}. 
The incident energy of photons was 60 keV, 
the scattering angle 160$^{o}$, and the momentum resolution of the 
experiment is estimated to be 0.12 a.u. The total number of accumulated 
counts under each directional profile was about $10^8$. The valence 
profiles of Al used for comparison were measured by Ohata 
{\it et al.} \cite{ohata} under the experimental conditions of the 
present measurements. However, we remeasured the [110] CPs of Al and 
Al$_{97}$Li$_3$ in the same experimental run in order to independently 
check the validity of the changes in the CPs with alloying discussed in 
this article.

Concerning computational details, selfconsistent electronic structures 
were first obtained in Al$_{1-x}$Li$_{x}$ at $x=0$ and $x=0.03$, 
respectively, using the KKR-LDA-CPA scheme. The total energy was not 
minimized, but instead the experimental lattice constant $a=7.6534$ a.u. 
for Al was used for both Al and the 3\% alloy. These results provided the 
input for evaluating the 3D momentum density $n({\bf p})$ in terms of 
the momentum matrix element of the KKR-CPA Green's function and the CPs 
for momentum transfer along the three high symmetry directions. The CPs 
for the limiting case of $x=0$ are in good accord with those obtained 
independently in Al via the KKR approach. The details of the KKR-LDA-CPA 
methodology and the associated momentum density and CP computations are 
described in Ref.~\cite{bansil_jpcs}. The accuracy of the 
theoretical CPs is estimated to be 1 part in $10^4$ so that changes in the 
CPs between Al and the 3\% Li alloy could be evaluated reliably.

\begin{figure} \begin{center}
\includegraphics[width=\hsize,width=7.0cm]{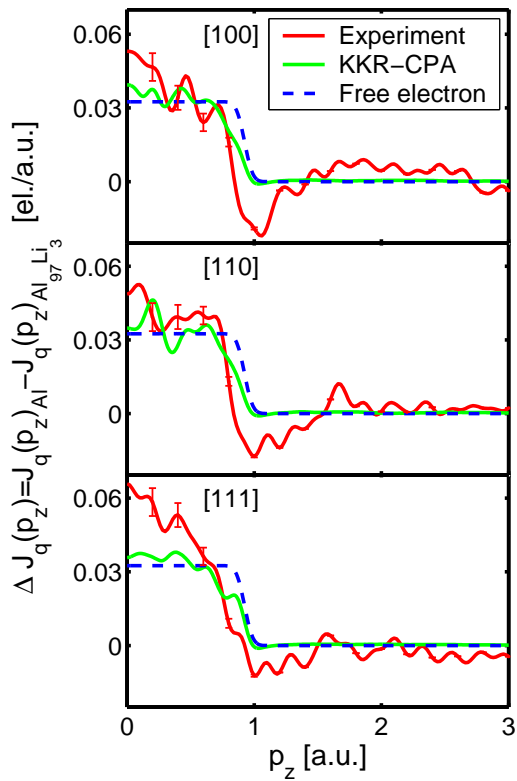}
\end{center}
\caption{
Experimentally observed changes $\Delta J_{\bf q}(p_{z})$
(Eq.~\ref{eq_dj})
between the valence 
[100], [110] and [111] 
CPs of Al and Al$_{97}$Li$_3$
are compared with
the corresponding results for
the \mbox{free-electron} model and the KKR-CPA theory.
}
\label{fig:001}
\end{figure}

Fig.~\ref{fig:001} considers measured and computed changes
\begin{equation}
\Delta J_{\bf q}(p_{z})=
J_{\bf q}(p_{z})_{Al}-J_{\bf q}(p_{z})_{Al_{97}Li_{3}},
\label{eq_dj}
\end{equation}
in the valence CP of Al for momentum transfer $p_z$ along the three
principal
symmetry directions when 3\% Li is added. $\Delta J_{\bf q}(p_{z})$
is a useful quantity because it allows us to focus on
the properties of only those electrons which are affected significantly by
alloying. We can obtain a handle on the size of
$\Delta J_{\bf q}(p_{z})$ by assuming all valence
electrons to be free electron like.
The momentum density $n({\bf p})$ is then uniform within the Fermi sphere
of radius $p_{F}$ and the corresponding CP is given
by the simple parabolic form
%\begin{equation}
%\label{equ:003}
$J_{\bf q}(p_{z})=
V_{WS}/(4\pi^2) (p_{F}^{2}-p_{z}^{2})\Theta(p_{F}-p_{z})$,
%\end{equation}
where $V_{WS}$ is the Wigner-Seitz cell volume and $\Theta$
is the step function.
The addition of 3 \% Li to Al causes a decrease of 0.06 valence
electrons/atom and a concomitant decrease in $p_{F}$ from 0.929 a.u. in Al
to 0.919 a.u. in Al$_{97}$Li$_{3}$. The use of these $p_{F}$ values
yields the isotropic free electron result
for $\Delta J_{\bf q}$ in Fig.~\ref{fig:001}
(dashed lines), with a value of $\approx 0.03$ extending to $p_z\approx 1$
a.u., which is representative of the scale of the experimentally observed
changes $\Delta J_{\bf q}(p_z)$.

The KKR-CPA curves in Fig.~\ref{fig:001} (red solid lines) are based on a 
realistic description of the electronic structures of both Al and 
Al$_{97}$Li$_3$ and display fine structure which reflects modifications of 
the free electron wavefunctions due to solid state effects. Nevertheless, 
it is striking that the KKR-CPA prediction differs significantly from the 
experimental results. This discrepancy is particularly noteworthy at and 
around the Fermi momentum of $\approx 1$ a.u. where the error bars on the 
experimental data are quite small. The data show a substantial negative 
excursion starting around $\approx$ 0.8 a.u. and non-zero values extending 
to high momenta. In sharp contrast, the KKR-CPA yields changes $\Delta 
J_{\bf q}(p_z)$, which are essentially zero beyond $\approx 1$ a.u. and 
positive in sign at all momenta. We have carried out extensive additional 
computations of the CP in a variety of ordered Al-Li crystal structures in 
order to ascertain if the aforementioned discrepancies in $\Delta J_{\bf 
q}$ may be related to clustering or ordering effects in the alloy which 
are missing in the KKR-CPA single site framework. In this connection, CPs 
were computed in Al$_3$Li and Al$_2$Li$_2$, but the $\Delta J_{\bf q}$'s 
so determined were similar to the KKR-CPA curves in all cases and in 
particular did not show a non-zero value for momenta greater than $\approx 
1$ a.u. It is clear that the discrepancies between theory and experiment 
in Fig.~\ref{fig:001} cannot be understood within the conventional 
LDA-based picture of the changes in the electronic spectrum of Al induced 
by the addition of a few percent Li impurities and indicate a failure of 
this picture at a fundamental level.

\begin{figure} \begin{center}
\includegraphics[width=\hsize,width=7.0cm]{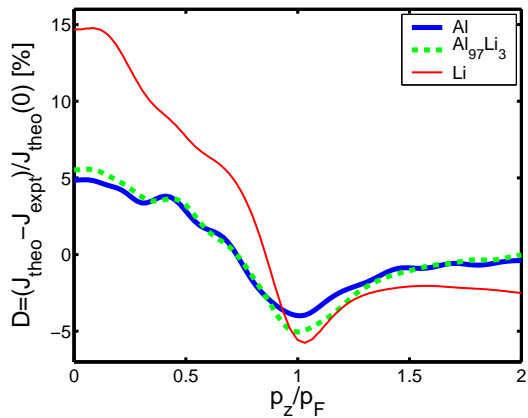}
\end{center}
\caption{
Normalized differences between the LDA-based theoretical
Compton profiles ($J_{theo}$) and the corresponding experimental profiles
($J_{expt}$) along the [100] direction in Li, Al and Al$_{97}$Li$_3$ as a
function
of the scaled momentum $p_z/p_F$, where $p_F$ is the Fermi momentum in
various
cases.}
\label{fig:002}
\end{figure}

Fig.~\ref{fig:002} further explores how the description of the ground 
state momentum density provided by the LDA breaks down in Al, Li and 
Al$_{97}$Li$_3$. Here the normalized deviation $D = (J_{theo} 
-J_{expt})/J_{theo}(0)$ between the theoretical and experimental CPs is 
considered (in percent) as a function of $p_z$ in units of $p_F$. In this 
way, the extent to which electron correlation effects beyond the LDA are 
at play in different systems can be identified and compared on a common 
scale. Although Fig.~\ref{fig:002} compares the CPs along the the [100] 
direction, results for other directional CPs are similar and are not shown 
for brevity. The deviations in Al (thick blue line) can be understood in 
terms of the Fermi liquid model where the break $Z_F$ in the momentum 
density at the Fermi energy, which is implicitly set equal to unity in the 
band theory calculations, is renormalized to a value of $0.7-0.8$ in line 
with the expected values in the correlated electron gas \cite{footnote2}. 
On the other hand, deviations in Li (thin red line) are seen to vary from 
+15\% at $p_z=0$ to -6\% around $p_F$ and are much larger than in Al 
\cite{agp}. The behavior of the momentum density of Li near the Fermi 
momentum has been investigated extensively in terms of CP experiments to 
adduce an effective $Z_F$ value of nearly zero \cite{schulke}, which is 
quite far from the corresponding electron gas predictions of 0.65-0.75 
\cite{footnote3}. Turning to Al$_{97}$Li$_3$ in 
Fig.~\ref{fig:002}, we see 
that the $D$-values in the alloy (dashed line) and Al differ by only a few 
percent at most momenta, but around the Fermi momentum, $D$-values in the 
alloy are comparable to those in Li. In other words, just a few percent Li 
atoms induce large changes in the electronic states of Al in a highly 
non-LDA like manner. The anomalous effects in the momentum density are 
seen to extend in the alloy over a substantial momentum range of $0.9\le 
p_z/p_F\le 1.4$ a.u. These observations are consistent with our earlier 
discussion on Fig.~\ref{fig:002} which also pointed to the presence of 
unusual discrepancies in $\Delta J_{\bf q}$ around $p_F$.

In order to gain insight into the nature of
discrepancy depicted in Fig.~\ref{fig:001},
we consider the spherical average
$\Delta J_{sph}$ of the change $\Delta J_{\bf q}$ in the valence CP
of Al due to alloying, which has
been discussed previously
in connection with Fig.~\ref{fig:001} above \cite{footnote4}.
$\Delta J_{sph}$ can be obtained in a cubic system via \cite{ohata}
\begin{equation}
\Delta J_{sph}=(10 \Delta J_{[100]}+16 \Delta J_{[110]}+
9 \Delta J_{[111]})/35.
\end{equation}
\begin{figure} \begin{center}
\includegraphics[width=\hsize,width=7.0cm]{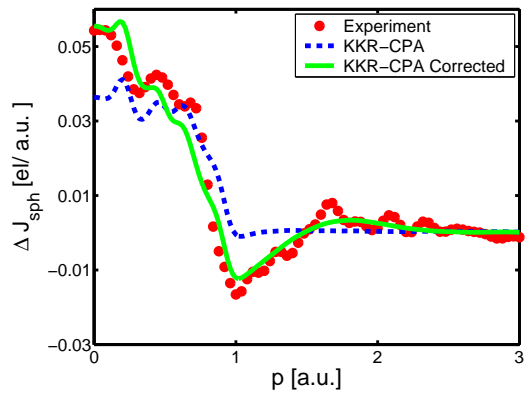}
\end{center}
\caption{
Spherical average $\Delta J_{sph}$ of the experimental $\Delta J_{\bf q}$
considered
in Fig.\ref{fig:001} is compared with the corresponding KKR-CPA result with
and
without the correction C(p) of Eq.~\ref{eq_c}.}
\label{fig:003}
\end{figure}

Fig.~\ref{fig:003} presents results for $\Delta J_{sph}$. As expected from 
the discussion of Figs.~\ref{fig:001} and~\ref{fig:002} above, the 
theoretical KKR-CPA curve (dashed line) differs substantially from the 
experimental data. The key to understanding this discrepancy in 
Fig.~\ref{fig:003} is to recognize that correlation effects can be viewed 
as exciting some electrons into higher energy unoccupied levels 
as a way of modifying the character of the ground state wavefunction.
The lowest unoccupied orbitals in Li and Al are the 2$p$ and the 
3$p$ orbitals, respectively. It is natural therefore to consider the 
effect of promoting an electron from an $s$ to a $p$ orbital. The 
resulting change in the CP is straightforwardly shown to be:
\begin{equation}
\nonumber
C(p)=w(
\left |\int R_0(r)
j_0(pr)r^2dr \right |^2-
\left |\int R_1(r)
j_1(pr)r^2dr \right|^2),
\label{eq_c}
\end{equation}
where $w$ is a weighting factor, $j_{\ell}$ are spherical Bessel 
functions, $R_{\ell}(r)$ are the radial wavefunctions in the crystal, and 
$\ell$ is the angular momentum index. The radial integrals extend to the 
Wigner-Seitz sphere radius $R_{WS}$. Specifically, we used Al 3$s$ and 
3$p$ radial wavefunctions in evaluating the right side of Eq.~\ref{eq_c}, 
but the shape of $C(p)$ is found to be essentially the same if Li 2$s$ and 
2$p$ wavefunctions are used instead \cite{footnote5}.

Fig.~\ref{fig:003} shows the result of correcting KKR-CPA via the term 
$C(p)$ where the weight factor $w$ of Eq.~\ref{eq_c} has been used as a 
fitting parameter to adjust the overall size of the correction term. A 
good fit is obtained for $w=0.03$/electron (solid green line). We 
emphasize that the deviations in the momentum density of pure Al compared 
to the LDA predictions, which were discussed above in connection with 
Fig.~\ref{fig:002}, can be explained reasonably within the 'standard' 
electron gas picture. Such standard correlation effects will of course 
also be present in Al$_{97}$Li$_3$ with a magnitude similar to that of Al 
because the electron density in Al and Al$_{97}$Li$_3$ is nearly the same 
with the $p_F$ values differing by only about 1\%. Therefore, when the 
difference $\Delta J_{\bf q}$ is formed in Eq.~\ref{eq_dj}, we would 
expect such standard electron gas type correlation effects to be 
cancelled. In other words, $w=0.03$/electron represents an {\it 
additional} correlation effect induced by the introduction of Li 
impurities in Al and implies that about 3\% valence electrons in 
Al$_{97}$Li$_3$ need to be promoted from $s-$ to $p-$ like states to 
account for the observed momentum density change in Al$_{97}$Li$_3$. We 
may look upon this result as indicating that correlations and the 
associated changes in the effective potentials produce an enhanced $p$ 
character of the ground state wave function of Al$_{97}$Li$_3$.  Note that 
since only 1\% of valence electrons in Al$_{97}$Li$_3$ come from Li atoms, 
the weight $w$ of 3\% cannot be explained in terms of Li electrons alone 
and indicates a significant involvement of Al electrons near the Li 
impurities in generating anomalous changes in the momentum density of the 
alloy.

In conclusion, our study provides direct evidence that the electronic 
ground state of Al is modified anomalously by the addition of Li 
impurities. The experimental momentum density in Al$_{97}$Li$_3$ is found 
to differ significantly from the predictions of the conventional 
Fermi-liquid type picture where correlation effects are included within 
the framework of the homogeneous electron gas. We show that the observed 
anomaly in the momentum density of the alloy can be accounted for if 3\% 
electrons in Al$_{97}$Li$_3$ are transferred from $s$- to $p$-like 
orbitals so that the $p$ character of the ground state wavefunction 
becomes enhanced. It is clear that Li impurities in Al constitute a 
relatively simple exemplar system in which the standard treatment of the 
interacting electron gas breaks down and properties of the correlated {\it 
inhomogeneous} electron gas must be considered.

%---------------------------------------------------------------------------
\acknowledgments
We thank Franek Maniawski for help with growing the single  
crystal of Al$_{97}$Li$_3$. 
This work was supported by the U.S.D.O.E. contract DE-AC03-76SF00098, by
the Polish Committee for Scientific Research, Grant Number 2 P03B 028 14
and benefited from the allocation of supercomputer time at NERSC and the
Northeastern University's Advanced Scientific Computation Center (ASCC).
This work was carried out with the approval of the Photon
Factory Advisory Committee, Proposal No. 97G288.

%----------------------------------------------------------------------------


\begin{thebibliography}{99}
%
% general alli
%
\bibitem {martin_review}
J.W. Martin, Ann. Rev. of Mat. Sci. {\bf 18}, 101 (1988).
%
%
%high visibility papers
%
%
\bibitem{ceder}
See e.g. A. Van der Ven and G. Ceder,
Phys. Rev. Lett. {\bf 94}, 045901 (2005).
%
% General Compton
%
\bibitem{cooper}
M.J. Cooper  {\em et al.} in {\em
X-Ray Compton Scattering}, Oxford University Press (2004).
\bibitem{kaplan}
I.G. Kaplan {\em et al.},
Phys. Rev. B {\bf 68}, 235104 (2003).
%
% KKR-CPA Alli
%
\bibitem{bansil_jpcs}
A. Bansil {\em et al.},
J. Phys. Chem. Solids {\bf 62}, 2191 (2001).
\bibitem{matsu}
I. Matsumoto {\em et al.},
Phys. Rev. B {\bf 64}, 045121 (2001).
\bibitem{suortti01}
P. Suortti {\em et al.},
J. Phys. Chem. Solids {\bf 62}, 2223 (2001).
\bibitem{matsu0}
I. Matsumoto {\em et al.},
J. Phys. Chem. Solids {\bf 61}, 375 (2000)
discuss a preliminary study 
of the [111] CP in Al$_{97}$Li$_3$.
%
% Method exp
%
\bibitem{sakurai92}
Y. Sakurai {\em et al.},
Rev. Sci. Instr. {\bf 63}, 1190 (1992).
\bibitem{ohata}
T. Ohata {\em et al.}, Phys. Rev. B {\bf 62}, 16528 (2000).
%
\bibitem{footnote2}
This is commonly referred to as the Lam-Platzman correction
where the momentum density obtained within the independent particle
framework of the band theory is corrected by using the momentum density of
the correlated homogeneous electron gas \cite{agp}.
%
% New paradigms
%
\bibitem{agp}
B. Barbiellini, A. Bansil, J. Phys. Chem. Solids {\bf 62}, 2181 (2001).
%
% Z_F in Li
%
\bibitem{schulke}  W. Sch\"ulke  {\em et al.},
Phys. Rev. B {\bf 54}, 14381 (1996).
%
\bibitem{footnote3}
Momentum density of Li cannot thus be
understood within the framework of correlated homogeneous electron
gas. Ref.~\cite{agp} considers the importance of the
correlated {\it inhomogeneous} electron gas in this connection.
%
\bibitem{footnote4}
There are of course smaller
anisotropic effects seen in Fig.~\ref{fig:003}, but their
quantitative nature is uncertain and we will not be concerned with these
effects here.
\bibitem{footnote5}
The radial wavefunctions in the crystal of
course depend on energy. In evaluating $C(p)$ we used wavefunctions at the
Fermi energy, although computations at other energies yielded little
change in the shape of $C(p)$.
%
\end{thebibliography}
\end{document}